\documentclass[onecolumn,amsmath,amssymb,prd,reprint,longbibliography,floatfix,8pt]{revtex4-1}

\usepackage{cancel}
\usepackage{enumitem}
\usepackage[utf8x]{inputenc}
\usepackage{graphicx}
\usepackage{dcolumn}
\usepackage{braket}
\usepackage{bm}
\usepackage[switch]{lineno}
\usepackage{float} 
\usepackage{subfigure}
\usepackage{tikz}
\usepackage{multirow}
\usetikzlibrary{arrows.meta}
\usetikzlibrary{arrows,decorations.pathmorphing,backgrounds,positioning,fit,petri}

\begin{document}

\title{Maximizing the magnetic anisotropy of Dy complexes by fine tuning organic ligands: A systematic multireference high-throughput exploration of over 30k molecules}

\author{Lion Frangoulis}
\author{Lorenzo A. Mariano}
\author{Vu Ha Anh Nguyen}
\author{Zahra Khatibi}
\author{Alessandro Lunghi}
\email{lunghia@tcd.ie}
\affiliation{School of Physics, AMBER and CRANN Institute, Trinity College, Dublin 2, Ireland}

\begin{abstract}
{\bf The design of the coordination environment of magnetic ions is key to achieving properties such as large magnetic anisotropy and slow magnetic relaxation, but a systematic exploration of the relevant chemical space for these compounds is missing. Here, we automatically extract all entries of mononuclear Dy coordination complexes from crystallographic databases and use multireference ab initio methods to compute their magnetic anisotropy. In addition, we generate and simulate magnetic anisotropy for 25k new molecules with the general formula [Dy(H$_2$O)$_5$L$_2$]$^{n-}$ and pentagonal bipyramidal coordination geometry, a motif selected as very promising. While no molecule with record magnetic anisotropy is serendipitously identified in crystallography databases, molecules with crystal field splittings over 1600 cm$^{-1}$ are identified by systematically exploring new organic ligands. This corresponds to a $\sim$100\% increase of magnetic anisotropy over the reference compound, $\sim$30\% over any known pentagonal bipyramidal Dy complex, and approaching record values of pseudo bi-coordinated Dy ions. This study demonstrates that the fine-tuning of Dy's second coordination sphere by organic ligands design can significantly improve magnetic anisotropy and that automated computational screening is key to accelerating this chemically non-intuitive process.}
\end{abstract}

\maketitle

\section*{Introduction}

The control of magnetization dynamics is central to many disciplines and technological applications, ranging from recording technologies\cite{Coey2010-qj} to magnetic resonance protein labels\cite{eaton2000relaxation} or contrast agents\cite{hingorani2015review}. Among all magnetic materials, magnetic molecules have played a central role in the development of our understanding of magnetism and its links to chemical and electronic structure \cite{coronado2020molecular}. In large part, this has been made possible by their synthetic versatility, which has allowed chemists to explore a large variety of spin topologies and couplings, and to establish their connection with magnetic behaviour \cite{sessoli2009strategies}. In this regard, Single-Molecule Magnets (SMMs) hold a special place. Besides having fascinated generations of scientists as a possible route towards nano-sized memory storage devices\cite{mannini2009magnetic} and spintronics\cite{bogani2008molecular} applications, SMMs have served as a benchmark to fully explore the phenomenology of magnetization dynamics in molecular systems as well as to probe the very limits of magnetic interactions in condensed matter systems\cite{zabala2021single}.

\begin{figure}[!h]
    \centering
    \includegraphics[width=\linewidth]{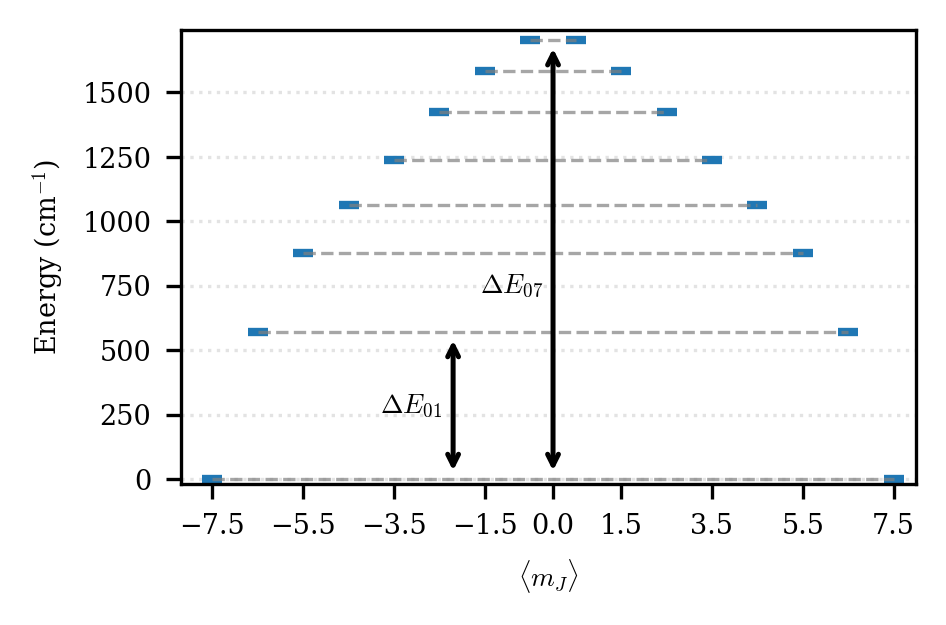}
    \caption{\textbf{Dysprosocenium energy levels.} The computed energy levels for [Dy(Cp$^{ttt}$)$_2$]$^{+}$ (Cp$^{ttt}$=C$_5$H$_2^t$Bu$_3$-1,2,4 and $^t$Bu=C(CH$_3$)$_3$) \cite{goodwin2017molecular} are plotted as a function of the reference $\hat{J}_z$ expectation value. Arrows indicate the energy gaps between the ground state and the first and last KDs.}
    \label{fig:energy_scheme}
\end{figure}

SMMs are essentially molecules characterized by a very slow magnetic relaxation, a phenomenon that finds its origin at the molecular level through a very specific electronic structure \cite{sessoli1993high}. For instance, in the case of Dy(III) coordination complexes, the most prominent example of SMMs, the ground-state multiplet $J=15/2$ of dysprosium is split in energy by crystal field interactions introduced by organic ligands. If such a splitting is, at the same time, very large and preserves the axiality of f-orbitals, then magnetic relaxation is often observed to become very slow\cite{rinehart2011exploiting}. Fig. \ref{fig:energy_scheme} illustrates how these conditions are fulfilled in a member of the dysprosocenium family, a group of notoriously slow-relaxing SMMs\cite{goodwin2017molecular,guo2018magnetic,mcclain2018high}, which shows 8 doubly degenerate states (Kramers Doublets - KDs) with almost pure $M_J$ values and large energy separation between ground state and both first ($\Delta E_{01}$) and 7th ($\Delta E_{07}$) excited KDs.

Our understanding of magnetization dynamics has eventually evolved beyond this simple picture, and we are now aware that magnetic relaxation is underpinned by spin-phonon transitions originated by the coupling of lattice vibrations and relativistic electronic interactions\cite{lunghi2022toward,mariano2025role,mondal2025spin}. However, despite the fact that several contributions to magnetic relaxation beyond the nature of the single ion's electronic structure affect magnetic relaxation (e.g. multi-ions exchange coupling\cite{demir2017giant,gould2022ultrahard}, the energy, density, and spin-phonon coupling of molecular vibrations\cite{lunghi2023spin}), the axiality and magnitude of the crystal field remain a key ingredient for SMMs and among the most important predictors for slow relaxation in Dy ions\cite{mondal2022unraveling}. Indeed, achieving such large axial splitting has been a central effort in molecular magnetism for the past 30 years\cite{zabala2021single}, and several synthetic design principles have been explored in the attempt to balance the proper magnetic anisotropy and chemical feasibility and stability. The work of Duan et al. \cite{duan2022data} provides an interesting overview of these efforts through a digital database (SIMDAVIS) and clearly summarizes the success of the strategy based on using strong axial donor atoms and weak, or the absence thereof, equatorial ligands to achieve the desired features of Dy's crystal field splitting. Further progress has occurred since the publication of SIMDAVIS\cite{duan2022data}, most notably with the observation of soft magnetic hysteresis up to 100 K\cite{emerson2025soft} and improved pentagonal bi-pyramidal SMMs\cite{luo2025supramolecular}, but the overall synthetic guidelines underpinning these latest achievements have remained qualitatively unchanged.

However, despite all these achievements, some fundamental questions remain wide open. When one looks at the actual number of mononuclear Dy complexes ever investigated, e.g. SIMDAVIS reports $\sim$650 entries for Dy, it becomes obvious that despite the high synthetic flexibility of coordination compounds, only a tiny portion of the chemical space of these molecules has likely been explored. This is particularly striking when these figures are compared with existing databases of organic molecules, e.g. the Cambridge Crystallography Structural Database (CCSD)\cite{groom2016cambridge}, including over one million entries, or the ZINC23 repository\cite{tingle2023zinc}, which includes billions of commercially available organic molecules. While this situation may cast shadows on our understanding of magneto-structural correlations in SMMs, it also points to a unique opportunity to search for novel molecules with record properties or new magnetic features entirely. In this regard, we note that while a good understanding is available for what concerns the relation between magnetic properties and first coordination shell structure\cite{mcadams2017molecular}, the effects of the second coordination shell are far less explored. In agreement with the common understanding that magnetic properties are extremely sensitive to small structural distortions, several reports suggest that the fine-tuning of ligands' secondary organic structure might play a key role in maximising magnetic properties\cite{goodwin2017molecular,guo2018magnetic,mcclain2018high} or even modulating spin-phonon coupling and magnetic relaxation\cite{luo2025supramolecular}, but a systematic exploration of these effects remains unattempted.

In this space, the use of ab initio simulations and models clearly plays an important role. In recent decades, computational methods have become a prominent tool for the investigation of (bio)chemical systems and materials and are rapidly evolving from a way to support experiments\cite{neese2019chemistry} to a fully-fledged driving force in the discovery process of novel compounds with target properties\cite{lunghi2022computational,choudhary2022recent}. High-throughput sampling\cite{curtarolo2013high} and machine learning techniques\cite{moosavi2020role} all played a significant role in this process and are rapidly shifting the paradigm of how molecular design is performed. While very promising, such methodologies are almost invariably developed and benchmarked over problems based on ground-state organic chemistry or inorganic materials\cite{sanchez2018inverse}. The design of coordination complexes and/or more challenging properties (e.g. magnetism and excited states) has only been preliminarily addressed, and several challenges still require a solution\cite{lunghi2022computational}. 

Here, we take on the challenge of accelerating the exploration of the chemical space for mononuclear Dy complexes and advance our understanding of magneto-structural correlations in SMMs prototypes. To achieve this, we build on the recently proposed approach to high-throughput multireference simulations \cite{mariano2024charting} and integrate it into a novel framework able to automatically explore the totality of the known chemical space for mononuclear Dy complexes, as well as to automatically generate novel ones. We perform a large-scale screening of about 30k Dy molecules and fully unravel the magneto-structural correlations underpinning large crystal field splitting in this class of SMMs. While the study of existing crystals confirms canonical design rules, the high-throughput generation of 25k molecules with pentagonal bi-pyramidal coordination makes the fine-tuning of the organic ligands beyond the connecting atoms emerge as a key handle to tune SMMs' properties to an unprecedented extent. A full analysis and discussion of these results unravels novel chemical strategies towards the optimization of SMMs and establishes advanced computational sampling techniques, such as high-throughput simulations, as a leading approach to the discovery of novel coordination complexes with tailored magnetic and electronic properties.

\section*{Results}

\begin{figure*}[t]
    \centering
    \includegraphics[width=\linewidth]{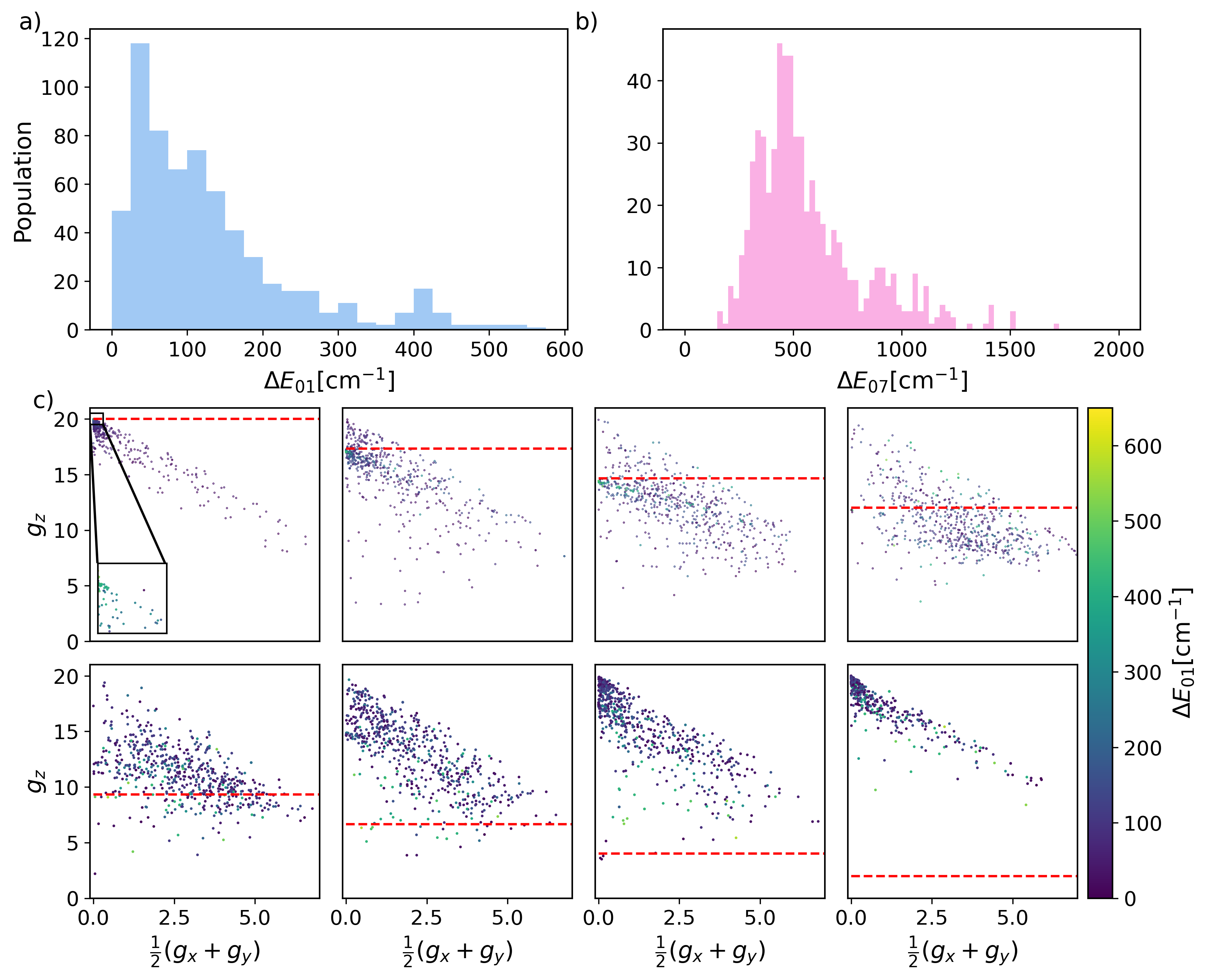}
    \caption{\textbf{Electronic properties of screened compounds.} Electronic properties of compounds deposited in the COD, CCSD, and SIMDAVIS databases, computed at the CASSCF level of theory. (a) Population distribution of the screened compounds with respect to the computed energy of the first electronic excited state. (b) Population distribution of the screened compounds with respect to the computed energy of the seventh electronic excited state, i.e., the highest electronic state of the fundamental multiplet $J = 15/2$. (c) Each panel shows the eigenvalues of the projected effective $S = 1/2$ $g$-tensor for the first eight KDs. The $g_z$ component is reported on the y-axis, while the x-axis represents the average value of the $g_x$ and $g_y$ components. The colour map of the points corresponds to the value of $\Delta E_{01}$ for each compound. The red dashed line shows the $g_z$ value for a perfect setup of axially aligned KD levels. }
    \label{fig:element_counts}
\end{figure*}

\textbf{Crystallography database screening.} Two essential steps are necessary to enable the automatic screening of the known chemical space of mononuclear Dy complexes: i) collect and determine all the corresponding three-dimensional crystallographic structures, and ii) design an automated, robust computational workflow to determine their magnetic properties with minimal user intervention. To achieve the first goal, we start by downloading all entries of the Crystallographic Open Database (COD)\cite{Vaitkus2021,Vaitkus2023,Merkys2023,Quiros2018,Grazulis2009,Grazulis2012,Grazulis2015,Downs2003}, the Cambridge Crystallography Structural Database (CCSD)\cite{groom2016cambridge}, and the SIMDAVIS database\cite{duan2022data}, which include at least one Dy atom (last datasets access in May 2023). For each crystal structure identified, we download the Crystallographic Information File (CIF) and process it to remove redundant atomic positions arising from static disorder. Specifically, we first use the disorder information stored in the CIF file to remove disordered atoms. We then convert the CIF to XYZ format using the Atomsk package\cite{Atomsk}, and verified the consistency between the experimental chemical formula and the resulting XYZ file. If any inconsistency is detected, the structure is discarded. Although this leads to the loss of some compounds, it greatly benefits the stability of the workflow. At this stage, the software MolForge\cite{lunghi2022toward} is used to identify and separate all the molecular units within the crystallographic unit cells, and polynuclear complexes are discarded. Finally, we retained only crystals that include a single type of mononuclear Dy complex (allowing for the presence of multiple molecules reported by symmetry). This last step is achieved by studying the fingerprints of every equivalent atom in the cell, as detailed in the Methods section.

The resulting dataset contains the structural information for neutral molecular crystals, from which the structure of each isolated magnetic molecule, together with its charge and multiplicity, must be extracted. To achieve this, we use a computational strategy based on multiple periodic Density Functional Theory (pDFT) calculations described in detail in the Methods section. In a nutshell, by knowing that the total charge of any unit cell is zero, we first perform a series of DFT calculations testing several plausible multiplicities for the Dy ion. The resulting Dy's local spin density is compared with the theoretically expected value, and multiplicities showing significant deviations are discarded. If a single multiplicity candidate remains, it is selected. Otherwise, in case of multiple acceptable results, the relative DFT energies are compared: when the energy difference exceeds a defined threshold, the lowest-energy solution is retained. If this last check also does not allow a clear determination of the ground-state multiplicity, the molecule is retained for further analysis (vide infra). The molecular charge is estimated using Hirshfeld and Mulliken population analyses. When the two approaches agree, their common value is assigned; otherwise, a range of intermediate charges is considered. If a unique combination of charge and multiplicity is identified at the DFT level, the information is stored. In cases where multiple candidates persist, each possible combination of charge and multiplicity is computed at the Complete Active Space Self-Consistent (CASSCF) level of theory, and the lowest energy solution is retained. 

After having determined the structure, charge, and spin multiplicity for each entry of this database, we perform multireference simulations with CASSCF on each molecule. The crystallographic atomic positions are used for this step. CASSCF calculations are then systematically used to determine the nature and energy of the low-lying states of Dy. CASSCF simulations are notoriously difficult to converge to the desired solution, and we here devise a new approach to achieve a stable and systematic convergence of these calculations. As a first step, we generate a suitable initial guess for the multiconfigurational wavefunction by performing a preliminary DFT calculation from which the natural molecular orbitals are used as input for the CASSCF procedure. A minimal active space is then systematically constructed using the seven 4f orbitals of the Dy ion and their corresponding electrons. Specifically,  we developed an automatic workflow that starts from the DFT set of input orbitals. The atomic composition of the input orbitals is analysed to ensure a dominant f-character, and orbital rotations are applied when necessary to enforce this condition. In practice, orbitals lacking sufficient f contribution (at least 30\%) are replaced by suitable candidates automatically selected from the inactive or virtual space, depending on their occupation, to recover a consistent 4f active space. This procedure is applied first to the input DFT orbitals and also to the subsequent converged CASSCF wavefunctions until no further rotation is required. If a proper convergence is not achieved with this method, a suitable set of molecular orbitals is constructed by combining DFT orbitals of the ligand framework (i.e., the molecule without the Dy ion) with those obtained from a CASSCF calculation performed on the isolated Dy(III) ion. In this way, the ionic character of the compound is preserved, and the initial orbitals are predominantly of f-character. 

\begin{figure*}
    \centering
    \includegraphics[width=1.0\linewidth]{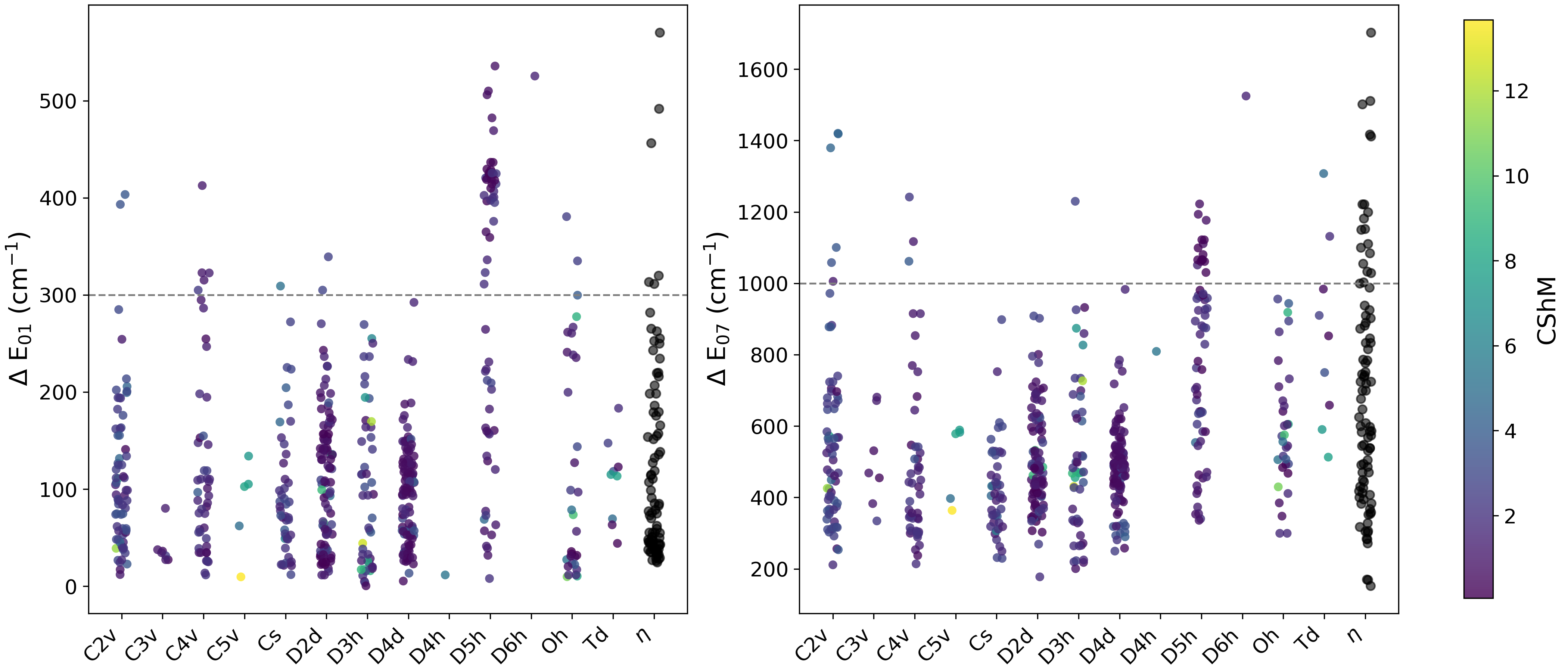}

    \caption{\textbf{First coordination shell symmetry analysis.} First ($\Delta E_{01}$, left) and last ($\Delta E_{07}$, right) excited KDs energies computed at the CASSCF level for each symmetry detected in the database. The color bar shows the CShM score of each compound, where smaller values indicate a closer match to the reference symmetry. Compounds containing hapticity-defined ligands bound to the magnetic center could not be assigned a reference symmetry with the SHAPE software and are shown as black dots and labeled as "$\eta$". }
    \label{fig:shape_database}
\end{figure*}

In total, we obtained the magnetic properties of 631 mononuclear Dy complexes with unique chemical compositions. Only 149 of these belong to the SIMDAVIS dataset, and only 265 are known as SMMs. The compounds span a wide range of molecular and crystal sizes, from 12 to 211 atoms per molecule, with an average of $\sim$85 atoms per structure, while crystals contain between 56 and 984 atoms with an average of 351 atoms. In the following, we first present the results of electronic structure simulations and then interpret them through magneto-structural correlations. 

The computed values of the excited KDs are reported in Fig. \ref{fig:element_counts}. The computed energy differences between the ground state and the first ($\Delta E_{01}$) and last ($\Delta E_{07}$) KDs of the fundamental multiplet $J=15/2$ reach up to 570.4 cm$^{-1}$ and 1703.0 cm$^{-1}$, respectively, corresponding to [Dy(Cp$^{ttt}$)$_2$]$^{+}$ (Cp$^{ttt}$=C$_5$H$_2^t$Bu$_3$-1,2,4 and $^t$Bu=C(CH$_3$)$_3$) \cite{goodwin2017molecular}. For reference, the molecule in the database with the largest $\Delta E_{01}$ that has not been reported as a SMMs is the pentagonal bipyramidal complex [Dy(H$_2$O)$_5$(Cy$_3$PO)$2$]$^{3+}$ \cite{lees2014complexes}, with $\Delta E_{01} = 426.1$ cm$^{-1}$ and $\Delta E_{07} = 1080.4$ cm$^{-1}$. Moreover, among the SMMs published after the SIMDAVIS database was collated, the pentagonal bipyramidal complex [Dy(L)$_2$(py)$_5$]$^+$ (HL = (S)-(-)-1-phenylethanol)\cite{yu2020enhancing} exhibits the largest value of $\Delta E_{01}$, computed at 535.7 cm$^{-1}$ and $\Delta E_{07} = 1176.5$ cm$^{-1}$. Overall, the mean values for $\Delta E_{01}$ and $\Delta E_{07}$ are 128.2 cm$^{-1}$ and 152.3 cm$^{-1}$, respectively. The large mean values of these figures reflect the anisotropic magnetic features of Ln and Dy ions in particular, which clearly exhibit a large crystal-field splitting compared to transition metal ions, even in the absence of molecular optimization. At the same time, the distributions reported in Fig. \ref{fig:shape_database}, also show how difficult it is to achieve very large values of these splittings, with very few molecular motifs populating the area of the plot for $\Delta E_{01} > 300$ cm$^{-1}$ or $\Delta E_{07} > 1000$ cm$^{-1}$. We now turn our attention to the values of the effective $g$-tensors computed by treating each KD as an effective spin-1/2. This analysis allows us to measure the axiality of these states, which would provide an ideal value of $g_z=g_J | M_J |$ in the case of perfect axiality, where $g_J = 4/3$ is Dy(III) Lande factor. The distribution of the computed effective $g$-tensor eigenvalues for each KD, reported in Fig. \ref{fig:element_counts}, shows that compounds with large $\Delta E_{01}$ also exhibit near-ideal values of $g_z$ for at least the first three KDs shown. For KDs of higher energies, the correlation is generally lost, pointing to an admixture among $M_J$ states due to a lack of perfect axial symmetry. 

To gain insights into the structure–property relationships within the selected dataset, we manually identify the first coordination sphere of each compound and analyze it systematically. Regarding the chemical composition of the first coordination sphere around the Dy ion, the most frequently occurring atoms are oxygen, nitrogen, and carbon, with average occurrences of 54.2\%, 27.5\%, and 13.4\%, respectively. Except for chlorine, which occurs at 2.8\%, all other elements occur at less than 1\%. The composition of the first coordination sphere is reported against the fundamental gap $\Delta E_{01}$ for different energy windows. Interestingly, it does not now show any specific trend, with the exception of confirming that the highest values of splitting are achieved in metalorganic compounds.  

We then perform the analysis of the first coordination shell geometry with the SHAPE software\cite{alemany2017continuous} to assign a reference symmetry. Specifically, the Continuous Shape Measures (CShM) method is employed to quantify the similarity between the atomic positions of each first coordination sphere and a set of reference polyhedral symmetries. A CShM value of zero indicates a perfect match with the reference symmetry, whereas larger values correspond to increasing deviations. For all complexes in the dataset, we computed CShM values for all symmetries compatible with the coordination number, and we selected the symmetry with the lowest score as the most representative of each structure. In Fig. \ref{fig:shape_database}, we plot $\Delta E_{01}$ and $\Delta E_{07}$ against the detected symmetries, with the colour map indicating the corresponding CShM values. The SHAPE software does not make it possible to assign a reference symmetry in the presence of haptic ligands ($\eta^5$, $\eta^6$, and $\eta^{8}$), and we report this class as one.

Interestingly, the distribution of computed $\Delta E_{01}$ values is significantly shifted toward larger splittings for compounds with $D_{5h}$ symmetry. Indeed, if we select $\Delta E_{01}=300$ cm$^{-1}$ as a threshold, we observe a total of 56 compounds as top-compounds according to this metric, and 37 out of 56 of these compounds possess $D_{5h}$ symmetry. Among the remaining molecules, 6 are bi-coordinated by $\eta$ ligands, 5 belong to $C_{4v}$ symmetry, 2 correspond to $C_{2v}$, $D_{2d}$ and $O_h$ symmetries, and 1 compound each exhibits $D_{6h}$ and $C_s$ symmetry. Interestingly, this distribution is not reproduced when $\Delta E_{07}$ is considered. A total of 47 compounds exhibit $\Delta E_{07}$ values larger than 1000 cm$^{-1}$, here taken as a reference for identifying top compounds. Among these, 19 contain $\eta$ ligands, 15 possess $D_{5h}$ symmetry, 6 to $C_{2v}$ symmetry, 3 belong to $C_{4v}$ symmetry,  2 to $T_d$ symmetry, and 1 compound each exhibits $D_{6h}$ and $D_{3h}$ symmetry. Similar conclusions were also reached by the developers of the SIMDAVIS database based on reported experimental effective reversal barriers\cite{duan2022data}. 

\begin{figure*}[t]
    \centering
    \includegraphics[width=\linewidth]{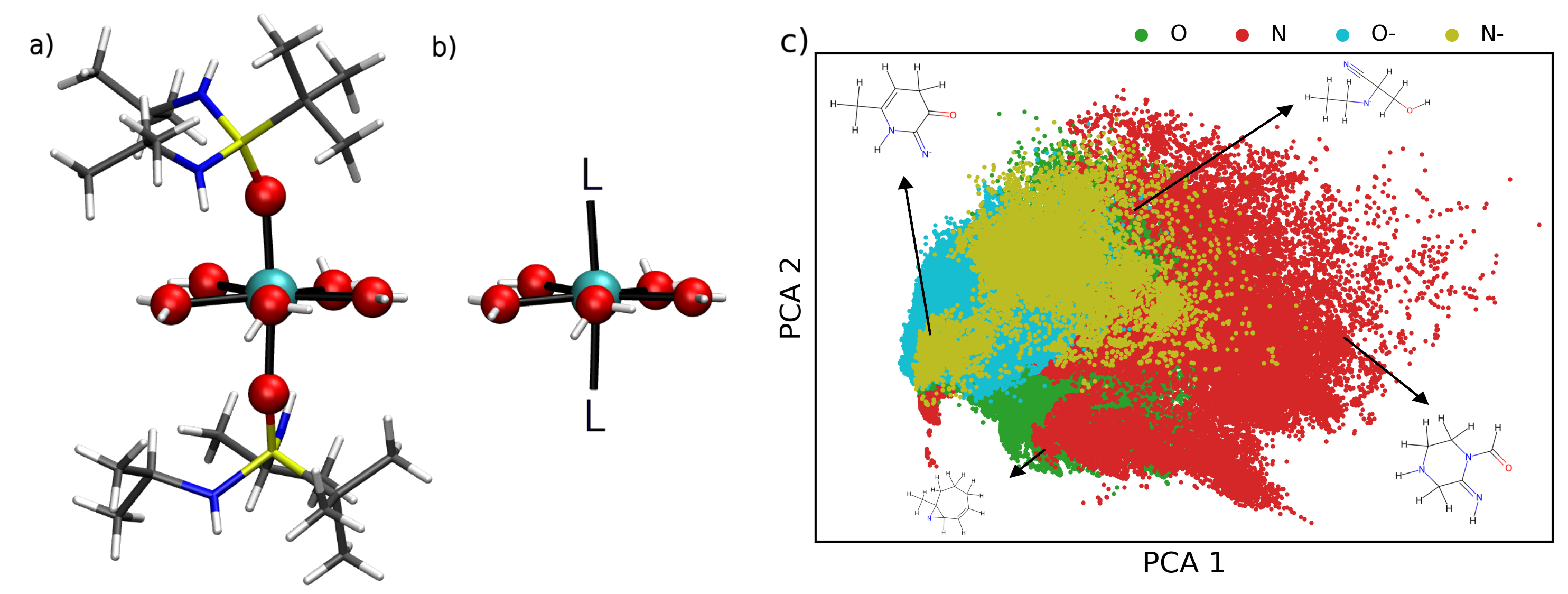}
    \caption{\textbf{Reference pentagonal bi-pyramidal compound and PCA.} Panel a) shows the three-dimensional structure of the compound [($^t$BuPO(NH$^i$Pr)$_2$)$_2$Dy(H$_2$O)$_5$]$^{3+}$\cite{Air-stable_2016}, where the central Dy(III) atom is surrounded by five water molecules in plane, and two $^t$BuPO(NH$^i$Pr)$_2$ ligands as the axial ligands. Panel b) shows the reduced structure of the Dy(III) core with the planar water ligands used to assemble new pentagonal bi-pyramidal compounds by inserting new axial ligands (L). Cyan: Dy, Red: O, Blue: N, Grey: C, White: H, Yellow: P. Panel c) shows the distribution of the principal components of the ligands' bispectrum components, sorted in colour by their different connecting atom species.}
    \label{fig:Original}
\end{figure*}

\textbf{Automatic generation of novel molecules.} While the previous analysis gives an overview of known compounds and helps in identifying the most successful trends yet to achieve large crystal-field splitting, it remains largely limited by i) the relatively small number of mononuclear Dy complexes available in existing literature, and ii) the underlying, unavoidable bias affecting their distribution, e.g. more compounds of some class might have been synthesized in excess because known for longer or because they are easier to make. In this second part of the work, we show how we can systematically explore novel portions of the chemical space of these compounds through high-throughput simulations. Having already established pentagonal bi-pyramidal geometry as one of the most favourable coordination motifs for Dy, we proceed to systematically replace the axial ligands of a reference compound with new ones in search of optimal values of crystal field splitting magnitude and axiality. For this purpose, we select the compound [($^t$BuPO(NH$^i$Pr)$_2$)$_2$Dy(H$_2$O)$_5$]$^{3+}$ \cite{Air-stable_2016}, depicted in Fig. \ref{fig:Original}a, as our benchmark model system. This molecule has a single Dy(III) ion with pentagonal bi-pyramidal geometry coordinated by five water molecules in plane and two $^t$BuPO(NH$^i$Pr)$_2$ ligands on the axial position, which confers an effective magnetic reversal energy barrier $U_{\rm eff}$ of 511.1 cm$^{-1}$. Importantly, this molecule was found to be air-stable, a rare yet necessary property for Dy-based SMMs' long-term applications. In addition, ab initio simulations have been previously used to characterize both static magnetic behaviour and magnetic relaxation of this benchmark system\cite{Air-stable_2016,mondal2022unraveling}, achieving good agreement with experiments, supporting the robustness of this new investigation. Finally, this compound has the advantage of a relatively small size due to small planar ligands, increasing the chance of accommodating axial ligands without incurring excessive steric repulsion.

The automatic creation of new compounds requires monodentate ligands of suitable charge, which we extracted from established databases of organic molecules. For this, we scanned through the 721k anions of the QM9star database\cite{QM9_Star_2024}. The latter is a dataset comprising approximately two million entries, including cations, anions, and radicals, along with their DFT-computed equilibrium structures and atomic information. The dataset also encompasses neutral organic molecules from the QM9 dataset \cite{QM9_2014}, from which the ions and radicals were derived by removing terminal hydrogens followed by DFT optimization. Using these coordinates and their charge states taken from the QM9star database, we extract molecules where the charged atom is a nitrogen or an oxygen, as well as the corresponding neutral molecules, and collate their XYZ files for a total of more than 200k organic ligands ready to act as axial ligands in the pentagonal bi-pyramidal Dy model compound. From this large set of ligands, we require an evenly sourced subset of manageable sizes for a high-throughput investigation. To achieve it, we produce an atomic fingerprint for each ligand's binding atom. The result is a 55-dimensional vector for each ligand, which gives a numerical representation of the local chemical environment. In order to simplify and visualize this data, we perform a Principal Components Analysis (PCA) and plot the two primary components in Fig. \ref{fig:Original}b, with each point representing one ligand. We randomly sample this distribution to gain 34146 samples. We then proceed to use these ligands to assemble novel prototypes of pentagonal bi-pyramidal Dy complexes following the schematics of Fig. \ref{fig:Original}b. As discussed in the \ref{sec:Methods} section, once a complex is assembled, its geometry is optimized with DFT (leading to 25044 optimized structures) and then used to run a CASSCF calculation to determine the crystal field splitting and axiality, resulting in a final 25024 compounds. 

\begin{figure*}[!htb]
    \centering
    \includegraphics[width=\linewidth]{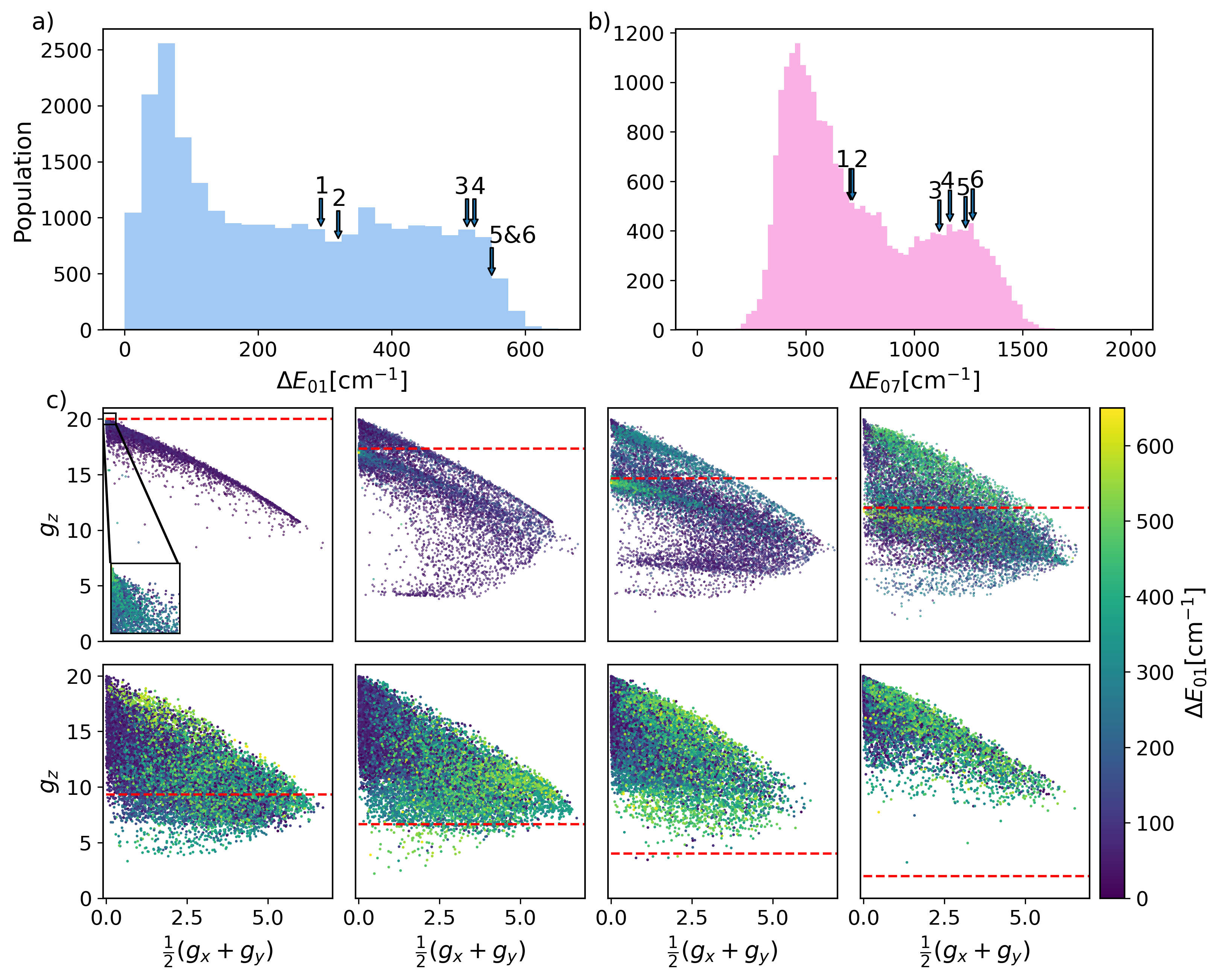}
    \caption{\textbf{Distribution of compounds in the database}. We first show the distribution of first (a) and last (b) KD energies compared to the best known reference compounds(Compound 1:\cite{Sourav_2022}, Compound 2: \cite{Air-stable_2016}, Compound 3-6: \cite{Qian-Cheng_2025}. Panel c) shows the relationship between the eight KDs $g_x, g_y$ and $g_z$ values, with their corresponding first KD gap visualized by colour. The inset zooms on the regime between 19.9 and 20 for $g_z$ and 0 and 0.01 for $g_x+g_y$. The red dashed line shows the $g_z$ value for a perfect setup of axially aligned KD levels.}
    \label{fig:dist}
\end{figure*}

\begin{figure*}[t]
    \centering
    \includegraphics[width=\linewidth]{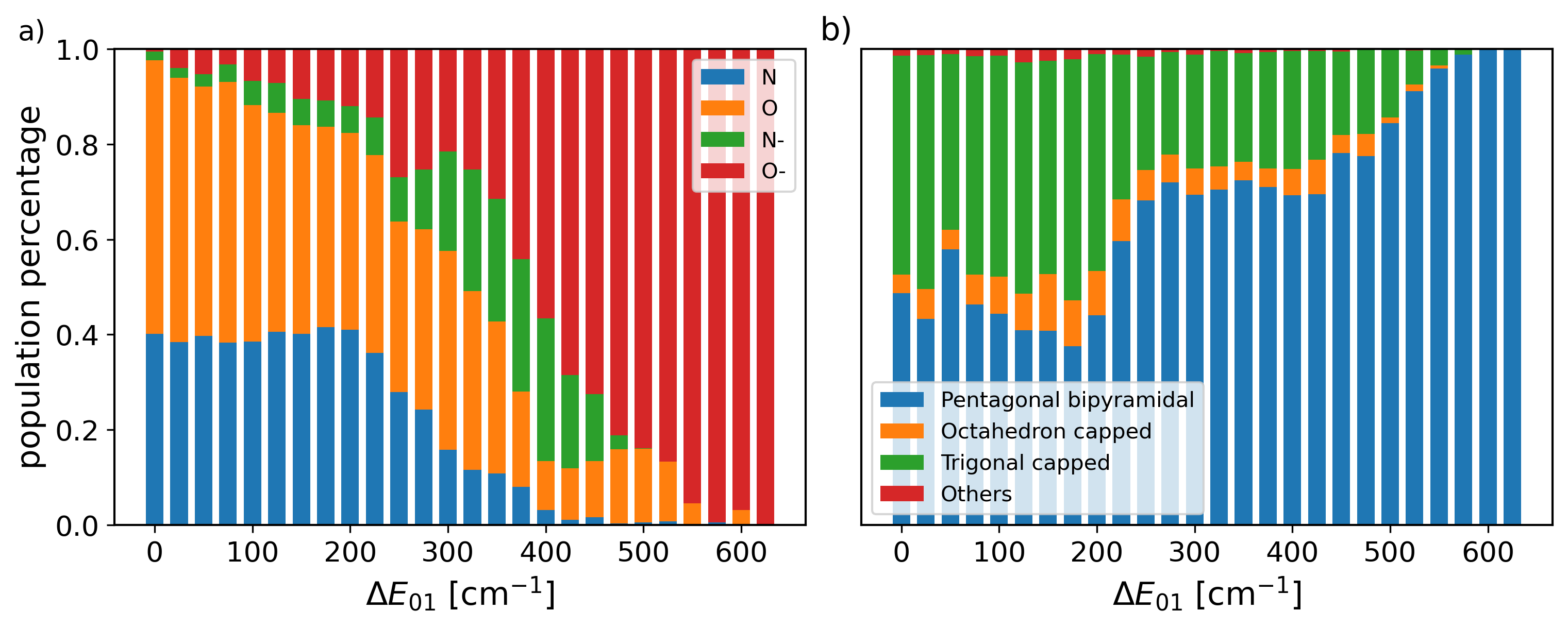}
    \caption{\textbf{Atomic species and geometry score distribution}. Panel a) shows the relative population of the axial ligands connecting atom species for different KD energy intervals. Panel b) shows the relative population of the most fitting coordination geometry symmetry for different KD energy intervals.}
    \label{fig:elem+shape}
\end{figure*}

Figs. \ref{fig:dist}a,b show $\Delta E_{01}$ and $\Delta E_{07}$ for this new dataset along with the corresponding metrics for 6 known SMMs, including two molecules with the same core as our molecular template (\textbf{1-2})\cite{Air-stable_2016,Sourav_2022}, and 4 among the top existing SMMs with $D_5h$ symmetry, each one presenting 5 pyridine equatorial ligands and diverse axial ones (\textbf{3-6})\cite{Qian-Cheng_2025}). Remarkably, different from the results of the screening of known molecules from crystallography databases, we here observe a large number of compounds populating the areas of the plot corresponding to $\Delta E_{01} > 300$ cm$^{-1}$ or $\Delta E_{07} > 1000$ cm$^{-1}$. Large average values of these metrics compared to those observed for crystallography databases are, in part, expected from the fact that we are biasing the sampling by pre-selecting molecules with a pentagonal bi-pyramidal geometry. However, the high-throughput study generates molecules exhibiting twice the crystal field splitting of the original reference compound, and the distribution of $\Delta E_{07}$ values well extend beyond the limits observed so far for pentagonal bi-pyramidal molecules. These results show that the tailoring of the organic axial ligand can have a dramatic effect on the magnetic anisotropy of Dy. Fig. \ref{fig:dist}c shows the distribution of $g_z$ versus the average of $g_x$ and $g_y$ for all eight KDs. The first KD shows a very clear inverse correlation between the axial and planar $g$-tensor. Moreover, the zoom around $g_z=20$ reveals that all high $\Delta E_{01}$ compounds cluster in that area. For excited KDs, the correlation becomes more complex, but it can be seen that high $\Delta E_{01}$ compounds are showing near ideal $g_z$ up to the fourth KD, indicating a crystal field simultaneously large in magnitude and axial in symmetry. 

We now turn to the analysis of the first coordination sphere, both in terms of chemical nature and symmetry. In Fig. \ref{fig:elem+shape} panel a), we analyse the chemical nature of the axial ligands' binding atom as a function of the metric $\Delta E_{01}$. Different from the same analysis for the crystallography databases, here we observe a clear preference for ligands with O$^{-}$ groups among the molecules with large crystal field splitting. This is in agreement with chemical intuition, which sees this chemical group as a strong donor, thus able to induce a large crystal field over the f electrons. Interestingly, while N$^-$ is also expected to be a strong donor, it is found to be less likely to lead to large crystal field splitting. We then perform the first-coordination shell symmetry analysis with the software SHAPE\cite{Shape}. Fig. \ref{fig:elem+shape} panel b) shows that several molecules do not preserve the initial $D_{5h}$ geometry and assume other coordination symmetries. This is clearly associated with a lowering of $\Delta E_{01}$, which is confirmed to be a good predictor for large crystal field splitting. 

\begin{figure*}[!htb]
    \centering
    \includegraphics[width=\linewidth]{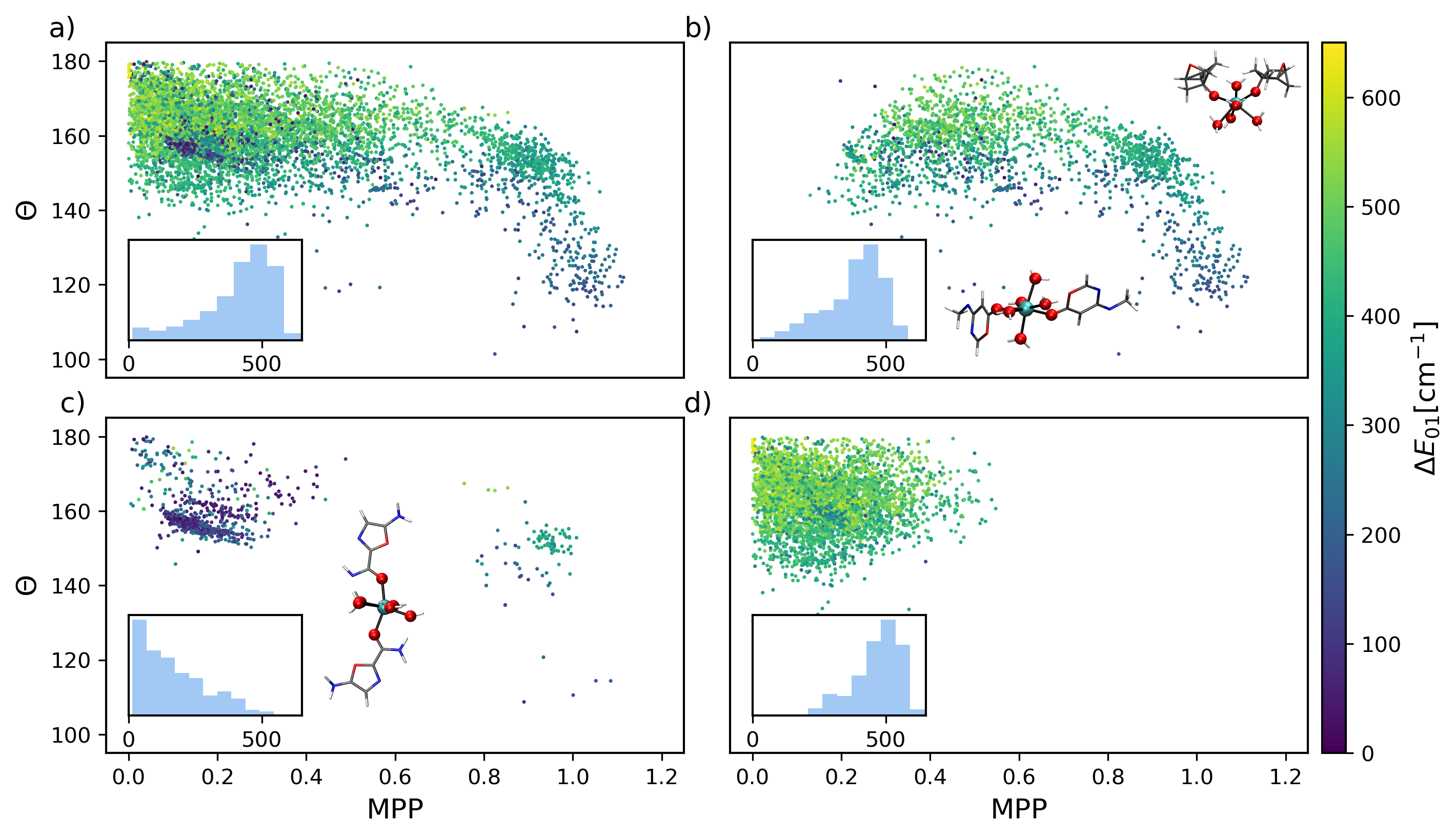}
    \caption{\textbf{Structural analysis}. Panel a) shows the total $\Theta$-MPP distribution of all O$^-$ compounds investigated and their KD energy in colour. Panel b) shows all compounds that are not classified as $D_{5h}$ by the SHAPE analysis. Panel c) shows all $D_{5h}$ compounds that have deviated from the original template structure. Panel d) shows all $D_{5h}$ compounds that have remained close to the original template structure.}
    \label{fig:geo}
\end{figure*}

Interestingly, but perhaps not surprisingly, this analysis clearly shows that pentagonal bi-pyramidal symmetry and O$^-$ strong axial donors are necessary but not sufficient conditions for achieving large crystal field splittings, pointing to fine structural details as the missing key to fully understand magneto-structural correlations for these molecules. We thus study the correlation between $\Delta E_{01}$ and two structural parameters: the angle $\Theta$ between the two axial connecting atoms, and the planarity of the five water molecules. The latter is captured by the MPP score\cite{MPP}, with a low MPP score corresponding to a closely planar arrangement of the water molecules. Fig. \ref{fig:geo}a reports the distribution of these two structural parameters vs $\Delta E_{01}$ for all molecules with O$^-$ ligands. No clear connection between $\Delta E_{01}$ and the structural parameters emerges at first glance. Indeed, we observe both large and small values of $\Delta E_{01}$ for molecules with both approximate $D_{5h}$ symmetry ($\Theta\sim 180^{\circ}$ and MPP $\sim 0$) and not. The case of molecules with symmetry different from $D_{5h}$, reported in Fig. \ref{fig:geo}b, is not hard to rationalize, as it will include both favourable and unfavourable coordination geometries. The presence of molecules with small $\Delta E_{01}$ and high $D_{5h}$ symmetry is instead explained by observing that two possible structural changes might occur during geometry optimization: i) one proton dissociates from a planar water molecule and binds the axial ligand, and ii) the ligands rearrange so that one water molecule takes the axial position of the pentagonal bi-pyramidal geometry. In both instances, the $D_{5h}$ nominal symmetry is preserved, but the local crystal field felt by Dy is drastically modified as reported in Fig. \ref{fig:geo}c. Finally, if only $D_{5h}$ molecules with a nominal structure similar to the prototype of Fig. \ref{fig:Original}b are retained, the distribution of Fig. \ref{fig:geo}d is obtained, confirming that $\Delta E_{01}$ correlates with how close the local geometry is to the ideal $D_{5h}$ one.

\begin{figure}[h!]
    \centering
    \includegraphics[width=\linewidth]{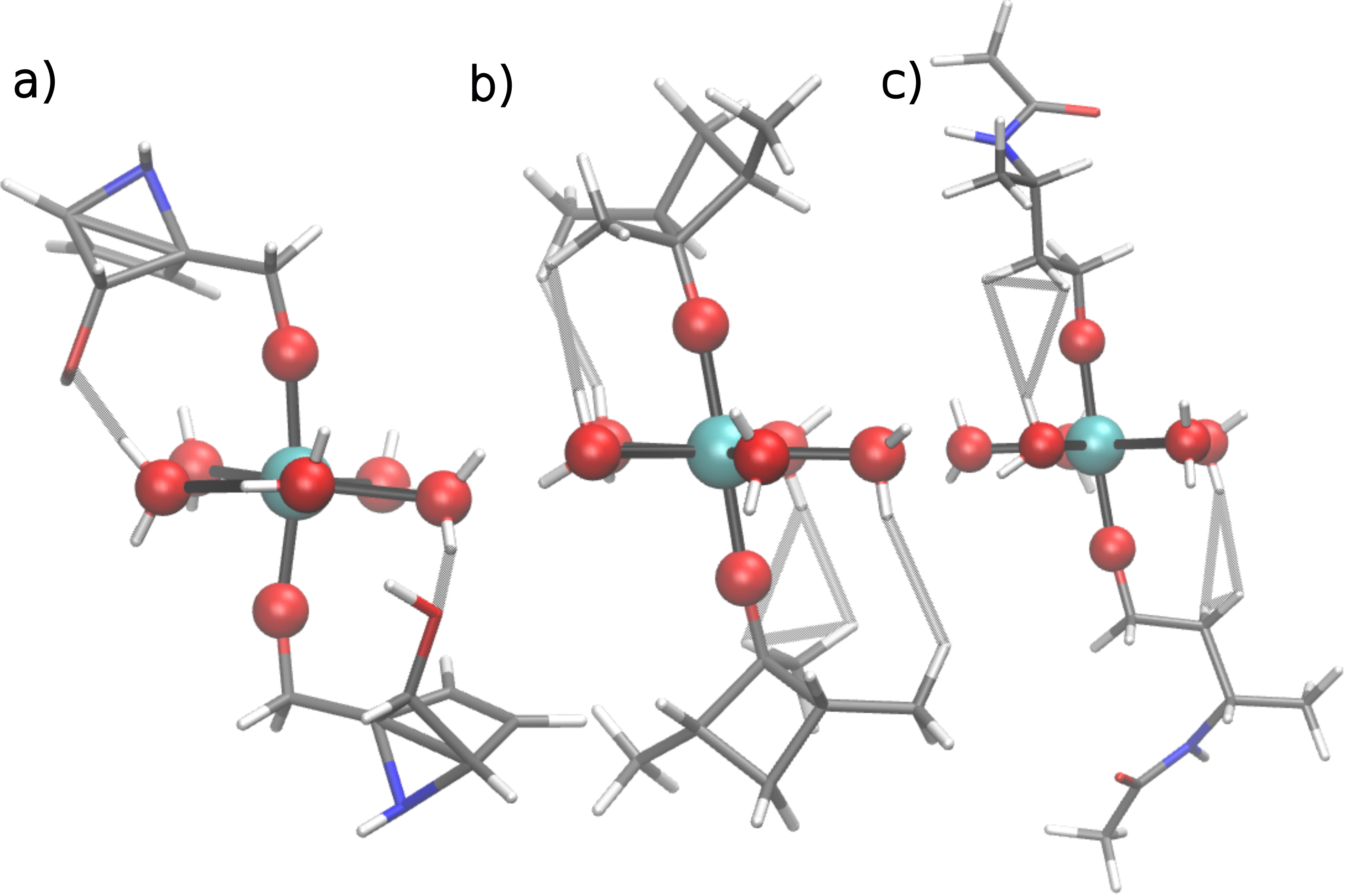}
    \caption{\textbf{$D_{5h}$ stabilization mechanism.} Panel a) shows the (O)-type interaction between planar water hydrogen atoms and an axial oxygen atom. Panel b) shows the (MC)-type of interaction between planar water hydrogen atoms and axial hydrogen atoms of multiple carbon atoms. Panel c) shows the (SC)-type of interaction between the hydrogen atoms of one carbon atom and the water hydrogens. Transparent bonds are used to highlight inter-ligand interactions.}
    \label{fig:Stabilization_Mechanisms}
\end{figure}

Finally, we perform a visual inspection of the top compounds, according to the metrics $\Delta E_{01}$ and $\Delta E_{07}$, with the goal of identifying which ligands' features, if any in particular, impart a nearly perfect $D_{5h}$ symmetry to the local Dy's geometry. Interestingly, we identify three distinct classes of molecules among those with $\Delta E_{01} > 600$ cm$^{-1}$. Each one presents a characteristic type of interaction between the water molecules' hydrogen atoms and the axial ligands' organic backbone. We group them into three different kinds of groups, based on the chemical nature of the groups involved in this interaction. The first group (O) includes compounds where such interaction is established with oxygen atoms (see Fig. \ref{fig:Stabilization_Mechanisms}a). The second group (MC) contains compounds where the hydrogen atoms of multiple axial carbon atoms each interact with the planar waters. The last category (SC) is composed of 
cases where the hydrogens from a single carbon atom, often belonging to a CH$_2$-CH$_2$ group following the binding atom, interact with the water's ones (see Fig. \ref{fig:Stabilization_Mechanisms}c). This last group contains all compounds above 630 cm$^{-1}$ and also the majority of high $\Delta E_{07}$ compounds. Table \ref{tab:stabilized_Kramers} reports energies and g-tensors for the three representative molecules displayed in Fig. \ref{fig:Stabilization_Mechanisms}, showing large energy gaps among all KDs and near ideal $g_z$ for the first four levels. 

\begin{table}[]
\footnotesize
\centering
\begin{tabular}{|c|c|c|c|c|c|c|c|c|c|}

\hline
Ligand & j&0&1&2&3&4&5&6&7\\
\hline
\multirow{4}{*}{a} &$\Delta E_{0j}$&0 &605 &1060 &1332 &1374 &1428 &1466 &1505 \\
 &$g_x$&0.00 &0.00 &0.05 &2.50 &1.54 &6.61 &2.95 &1.28 \\
 &$g_y$&0.00 &0.00 &0.04 &1.10 &0.72 &3.82 &1.04 &0.13 \\
 &$g_z$&19.99 &16.96 &14.20 &10.55 &16.14 &11.26 &15.36 &18.31 \\
\hline
\multirow{4}{*}{b} &$\Delta E_{0j}$&0 &624 &1110 &1426 &1526 &1548 &1566 &1639 \\
 &$g_x$&0.00 &0.00 &0.00 &0.04 &6.81 &3.51 &0.12 &0.08 \\
 &$g_y$&0.00 &0.00 &0.00 &0.03 &1.24 &1.61 &0.09 &0.01 \\
 &$g_z$&19.99 &17.02 &14.36 &12.13 &14.20 &9.09 &8.52 &17.41 \\
\hline
\multirow{4}{*}{c} &$\Delta E_{0j}$&0 &647 &1145 &1461 &1569 &1587 &1603 &1657 \\
 &$g_x$&0.00 &0.00 &0.00 &0.01 &9.31 &1.57 &0.78 &0.38 \\
 &$g_y$&0.00 &0.00 &0.00 &0.01 &1.36 &0.31 &0.62 &0.06 \\
 &$g_z$&19.99 &16.95 &14.23 &11.88 &11.91 &10.70 &8.26 &16.17 \\
\hline
Ideal & $g_z$&20&17.33&14.67&12&9.33&6.67&4&2\\ 
\hline

\end{tabular}
\caption{\textbf{Reference top compounds}. Energies and $g$-factors of all eight KDs of the best performing compounds (by their first Kramers energy) for each of the three stabilization mechanisms. The last row shows the $g_z$ factors of a perfectly axial Dy(III) compound for reference. Ligand names: a: 1219484\_80 (O), b: 829285\_87 (MC)r, c: 846701\_84 (SC).}
\label{tab:stabilized_Kramers}
\end{table}


\section*{Discussion and Conclusions}

While the exploration of the chemical space for organic molecules has been the target of intensive research\cite{dobson2004chemical}, the corresponding activity for coordination compounds, in particular for Ln and other magnetic ions\cite{lunghi2022computational}, is still in its infancy. Here, we have made a step towards closing this gap by designing a general computational workflow able to automatically screen crystallography databases as well as generate novel compounds starting from a selected molecular template. Importantly, this structure generation pipeline is linked to multireference calculations in a high-throughput fashion, allowing for accurate extraction of electronic and magnetic properties for this highly correlated class of molecules. From a methodological point of view, this result marks a potentially very significant step change in how these molecules could be designed. 

The screening of crystallography databases confirmed that the occurrence of highly anisotropic magnetic molecules is an extremely challenging task, and only very few viable molecular motifs have been found after decades of research on the topic. The reasons are multiple, and would likely include the challenging nature of the synthesis, the sensitivity of molecules to external conditions, and the need for time-consuming, multi-technique measurements to confirm the molecular structure and its properties. All these factors significantly slow down the progress rate in this field and, as a consequence, hamper the possibility of effectively exploring the chemical space underpinning this class of molecules. In this work, we have demonstrated that our automated computational approach cannot only rapidly screen the entire corpus of mononuclear Dy complexes ever crystallized, but it can produce orders of magnitude larger numbers of in-silico molecular prototypes. The introduction of such methods into the discovery pipeline of SMMs has the potential to drastically accelerate the discovery process, including extending it to more exotic coordination motifs that would be generally considered too risky and time-consuming endeavours. 

Clearly, the field of automated design of SMMs, and coordination compounds more in general, is just starting to get established, and several additional improvements will be necessary to fully and confidently integrate it within experimental discovery pipelines. We can identify three main directions that, in our opinion, could generate important progress in this regard: i) move towards a full coordination shell design, ii) include the design and generation of the entire crystal environment, and iii) account for molecular synthetic feasibility. For the present work, a detailed knowledge of successful coordination motifs was available from the study of crystallography databases and literature, but, looking forward to the deployment of this methodology to new magnetic properties or classes of molecules, the ability to access the entire chemical space will become essential. This is likely already within the reach of the present approach and would only require a corresponding scale-up in computational resources to target the wider chemical space that needs screening. On the other hand, points ii) and iii) represent significant challenges for which incremental solutions are not yet available. Some examples of molecular crystal design and synthetic probability scores exist for organic molecules\cite{price2014predicting,thakkar2021retrosynthetic,ertl2009estimation}, and while they represent interesting starting points, these methods do not easily apply to coordination compounds. For instance, coordination compounds often show a non-zero net charge and inherent structural flexibility, making the energy landscape of intermolecular interactions more complex than for rigid organic molecules and subject to the arbitrary choice of counterions and uncontrollable inclusion of ions and solvent in the crystal cell. Similarly, the synthesis of molecules like Dy SMMs is extremely sensitive to reaction conditions and often relies on uncharted reaction mechanisms, making stability or synthetic feasibility predictions dependent on too many variables. Notably, machine learning methods are making important progress on these challenges and hold promise to one day be able to manage these high levels of chemical complexity with computational workflows.

Importantly, regardless of the practical possibility of synthesizing the compounds generated in this study, the present results provide some novel and important insights into how ligand structure influences crystal field splitting in Dy complexes. A first general, important piece of evidence that emerges from the high-throughput study is the fact that secondary sphere structural perturbations can have a massive effect on Dy crystal field splitting. For instance, even if one just focuses on certified pentagonal bi-pyramidal molecules with O$^-$ donors in the axial position (inset of Fig. \ref{fig:geo}d), the distribution of $\Delta E_{01}$ spans a very large range of values of about 300 cm$^{-1}$. This result is particularly at odds with the conventional way to improve SMMs, which exclusively relies on changing the nature of the first coordination shell of Dy, either in terms of symmetry or binding atoms, and shows little regard to the exact nature of the second coordination shell. Our results not only suggest that the latter is a completely untapped resource in the optimization process, but it might be as efficient as the first coordination shell effects. If the present results are demonstrated to apply generally, and there is no reason to doubt they do, they could provide a significant boost to other existing coordination templates with even better prospects than the one picked for this study. 

The analysis of results also provides some initial chemical insights on how these inter-ligand interactions influence Dy's electronic structure, presenting the establishment of H$\cdots$H pairs among the water's protons and those belonging to the atoms beyond the binding one in the axial positions as the driving force for blocking the local coordination geometry very close to a perfect $D_{5h}$ symmetry. Although relatively weak, these kinds of interactions are known to play an important role in the stabilization of organic crystals and proteins and are ubiquitous in chemistry and biology\cite{matta2003hydrogen,matta2006extended}. This, in turn, suggests that the mechanism individuated within our high-throughput study is likely very general and can find widespread application beyond the specific equatorial ligands employed here and the case of $D_{5h}$ symmetry. Interestingly, a recent study has employed evolutionary optimization to the same molecular template used here and have individuated an alternative stabilization mechanism\cite{frangoulis2025generating}, based on generating ligands with side chains containing electronegative groups able to interact with the planar waters' protons. Although the specific chemical stabilization mechanism is different from the one observed here, they appear to fall under the same general concept, supporting our claim on the general importance that supramolecular interaction might play in indirectly shaping the electronic properties of coordination complexes. Moreover, once the importance of supramolecular interactions is recognized, several mechanisms for exploiting this design degree of freedom may come to mind. These include the possibility to control molecular structure through intermolecular interactions, where the crystal lattice can be reimagined as an active element shaping molecular properties instead of just an inert environment. Taking a step further, the recent studies of Luo et al. and Swain et al. have shown the importance of supramolecular interactions in shaping spin-phonon relaxation\cite{luo2025supramolecular,swain2025encapsulation}, giving a glance at the potential of crystal engineering for shaping both static and dynamic magnetic properties of SMMs. 

Last but not least, we would like to remark that the framework proposed here is not limited to SMMs or Dy ions, and it can readily be extended to other classes of coordination compounds. In particular, the integration of multireference methods into the high-throughput methodology greatly expands the remit of these computational techniques to properties that often remained beyond the limits of what is achievable with conventional electronic structure theory, e.g. density functional theory. Magnetic and excited states properties are two topical examples with very high technological relevance that can be tackled with this approach.

In conclusion, we have performed a systematic computational screening of all known mononuclear complexes of Dy and determined the most successful trends to achieve large crystal field splitting values. In addition, we generated 25k new molecules based on a promising pentagonal bi-pyramidal template and demonstrated that the modulation of the interaction among atoms in the second coordination sphere can play a crucial role in maximising crystal field splitting. On the one hand, these results introduce new conceptual tools to think about the chemical design of SMMs, and more generally of coordination compounds with tailored magnetic and optical properties. On the other hand, they illustrate the power of high-throughput simulations in the determination of optimal ligands structures where the fine effects of supramolecular interactions might fail simple chemical intuition.

\section*{Computational Methods}
\label{sec:Methods}
\textbf{QM9 database extraction.} From the extracted set of molecules, we then removed the negative charge and compared the resulting neutral molecule with neutrals in the QM9star. Those that were present in the QM9star neutral set, together with the corresponding anions were then extracted to build the original dataset for the high-throughput study. Before compiling and generating the full molecules using the compounds in the dataset, it is important to note that the connecting head of the ligand is an essential element in assembling the full compounds using standard techniques. The connecting head is fixed to the charged atom in the anions, however, it could range between an arbitrarily choices of atoms in neutrals. For this reason, we first scanned through the neutral ligands and set the connecting atom to be any of the nitrogen or oxygen present in the ligand. This led to multiple sets of neutral ligands from a unique ligand.

\textbf{Automatic compound creation pipeline.} The Dy(III) are created using a core of Dy(H$_2$O)$_5$, pre-optimized with respect to the distances between atoms, but restricted to perfect pentagonal symmetry. Using the software MolSimplify \cite{molSimplify}, two copies of the ligand are attached in a linear coordination around this pentagonal core, creating a pentagonal bi-pyramidal structure. Following this step, the structure is optimized using DFT in two steps: first under constraints that enforce the planar oxygen atoms to stay within a perfect plane and pentagonal symmetry, allowing the ligand to optimize internally in the vicinity of the Dy ion, with minimum disturbance from the axial ligands. Once this step is converged, another optimization cycle is started without any constraints, leading to a final geometry, which is then used for the extraction of magnetic parameters using CASSCF and structural parameters. 

\textbf{Electronic structure calculations}
We use the quantum chemistry software ORCA 5 \cite{neese2020orca} to perform both gas-phase DFT and state-averaged CASSCF calculations. Scalar relativistic effects are treated using the Douglas–Kroll–Hess (DKH) method, with picture-change effects included up to second order to account for DKH corrections in the spin–orbit coupling operator. The DKH-def2-TZVPP basis set is employed for all atoms except those heavier than Kr, for which the SARC-DKH-TZVPP basis set is used. Energies and forces are converged to $10^{-9}$ a.u. and $10^{-6}$ a.u., respectively. Due to the convergence issues of DFT on lanthanides, we replace the Dy(III) core with a Y(III) core for the geometry optimization. Their electronic structure is highly similar except for the lack of f-like orbitals, which are highly localized on the core itself and contribute negligibly to the bonding behaviour. Once the geometry optimization is converged, the core is then again replaced with Dy(III) for the CASSCF calculation. The active space contains 9 electrons in 7 orbitals, with 21 roots and a final multiplicity of 6. The CASSCF calculation returns both the KDs energies and the effective $g$-tensors. DFT calculations performed on the periodic unit cells of the database compounds were performed using CP2K \cite{kuhne2020cp2k} with the PBE functional, a plane wave cutoff of 1200 Ry, and the DZVP-MOLOPT-SR-GTH basis set for all atoms. 

\textbf{Structural parameter evaluation.} In order to differentiate different ligands and analyse the chemical space available from the database, we first want to create a quantity describing the similarity of different ligands, and opt to use bispectrum components \cite{Bartok_2013} using the software MolForge \cite{lunghi2022toward}. We chose a cut-off distance of 4\AA \ around the ligand's connecting atom, and order $2J=8$. This results in a 55-dimensional vector describing the environment around the connecting atom. The molecular planarity score (MPP)\cite{MPP} is defined by fitting a plane through the centre of mass of the five planar oxygen atoms using a least square fitting, and calculates the distances from atom $i$ to this plane $d_i$, and their maximum and minimum $d_{max}$ and $d_{min}$, to calculate the molecular planarity score as
\begin{align}
    \text{MPP}=\sqrt{\frac{1}{5}\sum_{i=1}^5d_i^2}\:.
\end{align}
This score is also used to automatically detect ligand swapping: we calculate the score not only for the five planar oxygen atoms, but also for combinations of planar and axial connecting atoms. If the score for any of these combinations is lower than the original planar one, there is a combination that creates a more aligned plane, indicating that the axial and planar ligands have swapped position around the Dy ion.\\

\vspace{0.2cm}
\noindent
\textbf{Acknowledgements and Funding}\\
This project has received funding from the European Research Council (ERC) under the European Union’s Horizon 2020 research and innovation programme (grant agreement No. [948493]). Computational resources were provided by the Trinity College Research IT and the Irish Centre for High-End Computing (ICHEC) and EURO-HPC.

\vspace{0.2cm}
\noindent
\textbf{Authors Contributions}\\
L.F. and L.A.M contributed equally to the work. A.L. proposed the project and supervised the work. NVHA designed the high-throughput scheme for crystallography databases. L.A.M carried out the ab initio simulations of the crystallography database and its analysis. L.F. carried out the high-throughput simulation of the generated molecules and analyzed the results with L.A.M. Z.K. extracted and curated the organic ligands database. All authors contributed to the discussion of the results. L.F., L.A.M and A.L. wrote the manuscript with input from all authors.

\vspace{0.2cm}
\noindent
\textbf{Data Availability}\\
The structural data and magnetic properties of all compounds analysed in this work are available under \url{https://doi.org/10.5281/zenodo.19387957}, and are split into two datasets. One contains the compounds extracted from the COD, CCSD and SIMDAVIS database, their structure in .xyz file format and their g-tensors and energy separations of all eight Kramers levels. The second contains the same information for all 25024 compounds generated in the high-throughput study.

\vspace{0.2cm}
\noindent
\textbf{Conflict of interests}\\
The authors declare that they have no competing interests.

\bibliographystyle{naturemag}
\bibliography{refs}

@Article{QM9_Star_2024,
author={Tang, Miao-Jiong
and Zhu, Tian-Cheng
and Zhang, Shuo-Qing
and Hong, Xin},
title={QM9star, two Million DFT-computed Equilibrium Structures for Ions and Radicals with Atomic Information},
journal={Scientific Data},
year={2024},
month={Oct},
day={21},
volume={11},
number={1},
pages={1158},
doi={10.1038/s41597-024-03933-6},
}

@misc{QM9_2014, 
title={Quantum chemistry structures and properties of 134 kilo molecules}, 
url={https://springernature.figshare.com/collections/Quantum_chemistry_structures_and_properties_of_134_kilo_molecules/978904/5}, 
DOI={10.6084/m9.figshare.c.978904.v5}, publisher={figshare}, 
author={Ramakrishnan, Raghunathan and Dral, Pavlo and Rupp, Matthias and Anatole von Lilienfeld, O.}, year={2014}, 
month={Jul} }

@article{emerson2025soft,
  title={Soft magnetic hysteresis in a dysprosium amide--alkene complex up to 100 kelvin},
  journal={Nature},
  volume={643},
  number={8070},
  pages={125--129},
  year={2025},
  publisher={Nature Publishing Group UK London}
}

@article{duan2022data,
  title={Data-driven design of molecular nanomagnets},
  author={Duan, Yan and Rosaleny, Lorena E and Coutinho, Joana T and Gim{\'e}nez-Santamarina, Silvia and Scheie, Allen and Baldov{\'\i}, Jos{\'e} J and Cardona-Serra, Salvador and Gaita-Ari{\~n}o, Alejandro},
  journal={Nature Communications},
  volume={13},
  number={1},
  pages={7626},
  year={2022},
  publisher={Nature Publishing Group UK London}
}

@article{Shape,
author = {Tuvi-Arad, Inbal and Shalit, Yaffa and Alon, Gil},
title = {CSM Software: Continuous Symmetry and Chirality Measures for Quantitative Structural Analysis},
journal = {Journal of Chemical Information and Modeling},
volume = {64},
number = {14},
pages = {5375-5380},
year = {2024},
doi = {10.1021/acs.jcim.4c00609},
note ={PMID: 38954801},

}

@article{Sourav_2022,
author = {Gupta, Sandeep K. and Dey, Sourav and Rajeshkumar, Thayalan and Rajaraman, Gopalan and Murugavel, Ramaswamy},
title = {Deciphering the Role of Anions and Secondary Coordination Sphere in Tuning Anisotropy in Dy(III) Air-Stable D5h SIMs},
journal = {Chemistry – A European Journal},
volume = {28},
number = {4},
pages = {e202103585},
doi = {https://doi.org/10.1002/chem.202103585},
year = {2022}
}

@Article{Air-stable_2016,
author ="Gupta, Sandeep K. and Rajeshkumar, Thayalan and Rajaraman, Gopalan and Murugavel, Ramaswamy",
title  ="An air-stable Dy(iii) single-ion magnet with high anisotropy barrier and blocking temperature",
journal  ="Chem. Sci.",
year  ="2016",
volume  ="7",
issue  ="8",
pages  ="5181-5191",
publisher  ="The Royal Society of Chemistry",
doi  ="10.1039/C6SC00279J",
}

@article{Qian-Cheng_2025,
author = {Qian-Cheng Luo  and Lorenzo A. Mariano  and Wen-Jie Xu  and Zi-Han Li  and Ke-Xin Yu  and Yan Peng  and Alessandro Lunghi  and Yan-Zhen Zheng },
title = {Supramolecular Interactions Modulate Raman Relaxation in D5h Symmetric Dy(III) Single-Molecule Magnets Opening the Magnetic Hysteresis up to 50 K},
journal = {ChemRxiv},
volume = {2025},
number = {0731},
pages = {},
year = {2025},
doi = {10.26434/chemrxiv-2025-jrlh0},
}

@Article{Vaitkus2023,
  author    = {Vaitkus, Antanas and Merkys, Andrius and Sander, Thomas and Quirós, Miguel and Thiessen, Paul A. and Bolton, Evan E. and Gražulis, Saulius},
  title     = {A workflow for deriving chemical entities from crystallographic data and its application to the {C}rystallography {O}pen {D}atabase},
  journal   = {Journal of Cheminformatics},
  year      = {2023},
  volume    = {15},
  number    = {1},
  month     = {Dec},
  doi       = {10.1186/s13321-023-00780-2},
  publisher = {Springer Science and Business Media LLC},
}

@Article{Merkys2023,
  author    = {Merkys, Andrius and Vaitkus, Antanas and Grybauskas, Algirdas and Konovalovas, Aleksandras and Quir{\'{o}}s, Miguel and Gra{\v{z}}ulis, Saulius},
  title     = {Graph isomorphism-based algorithm for cross-checking chemical and crystallographic descriptions},
  volume    = {15},
  doi       = {10.1186/s13321-023-00692-1},
  number    = {1},
  journal   = {Journal of Cheminformatics},
  publisher = {Springer Science and Business Media LLC},
  year      = {2023},
  month     = {Feb},
}

@Article{Vaitkus2021,
  author   = {Vaitkus, Antanas and Merkys, Andrius and Gražulis, Saulius},
  title    = {Validation of the {C}rystallography {O}pen {D}atabase using the {C}rystallographic {I}nformation {F}ramework},
  journal  = {Journal of Applied Crystallography},
  year     = {2021},
  volume   = {54},
  number   = {2},
  pages    = {661--672},
  month    = {Apr},
  doi      = {10.1107/S1600576720016532},
}

@Article{Quiros2018,
  author    = {Miguel Quir{\'{o}}s and Saulius Gra{\v{z}}ulis and Saul{\.{e}} Girdzijauskait{\.{e}} and Andrius Merkys and Antanas Vaitkus},
  title     = {Using {SMILES} strings for the description of chemical connectivity in the {C}rystallography {O}pen {D}atabase},
  journal   = {Journal of Cheminformatics},
  year      = {2018},
  volume    = {10},
  number    = {1},
  month     = {May},
  doi       = {10.1186/s13321-018-0279-6},
  publisher = {Springer Nature},
}

@ARTICLE{Grazulis2015,
  author = {Gražulis, Saulius and Merkys, Andrius and Vaitkus, Antanas and Okulič-Kazarinas,
            Mykolas},
  journal = {Journal of Applied Crystallography},
  pages = {85-91},
  title = {Computing stoichiometric molecular composition from crystal structures},
  year = {2015},
  month = {Feb},
  number = {1},
  volume = {48},
  doi = {10.1107/S1600576714025904},
}

@article{Grazulis2012,
author = {Gražulis, Saulius and Daškevič, Adriana and Merkys, Andrius and Chateigner, Daniel and Lutterotti, Luca and Quirós, Miguel and Serebryanaya, Nadezhda R. and Moeck, Peter and Downs, Robert T. and Le Bail, Armel}, 
title = {Crystallography Open Database (COD): an open-access collection of crystal structures and platform for world-wide collaboration}, 
volume = {40}, 
number = {D1}, 
pages = {D420-D427}, 
year = {2012}, 
doi = {10.1093/nar/gkr900},  
journal = {Nucleic Acids Research} 
}

@article{Grazulis2009,
author = "Gra{\v{z}}ulis, Saulius and Chateigner, Daniel and Downs, Robert T. and Yokochi, A. F. T. and Quir{\'{o}}s, Miguel and Lutterotti, Luca and Manakova, Elena and Butkus, Justas and Moeck, Peter and Le Bail, Armel",
title = "{Crystallography Open Database {--} an open-access collection of crystal structures}",
journal = "Journal of Applied Crystallography",
year = "2009",
volume = "42",
number = "4",
pages = "726--729",
month = "Aug",
doi = {10.1107/S0021889809016690},
}

@ARTICLE{Downs2003,
  author = {Downs, R. T. and Hall-Wallace, M.},
  title = {The American Mineralogist Crystal Structure Database},
  journal = {American Mineralogist},
  year = {2003},
  volume = {88},
  pages = {247-250},
}

@article{groom2016cambridge,
  title={The Cambridge structural database},
  author={Groom, Colin R and Bruno, Ian J and Lightfoot, Matthew P and Ward, Suzanna C},
  journal={Structural Science},
  volume={72},
  number={2},
  pages={171--179},
  year={2016},
  publisher={International Union of Crystallography}
}

@article{yu2020enhancing,
  title={Enhancing magnetic hysteresis in single-molecule magnets by ligand functionalization},
  author={Yu, Ke-Xin and Kragskow, Jon GC and Ding, You-Song and Zhai, Yuan-Qi and Reta, Daniel and Chilton, Nicholas F and Zheng, Yan-Zhen},
  journal={Chem},
  volume={6},
  number={7},
  pages={1777--1793},
  year={2020},
  publisher={Elsevier}
}

@article{lees2014complexes,
  title={Complexes of lanthanide chlorides with tricyclohexylphosphine oxide. The single crystal X-ray structures and solution properties of pentagonal bipyramidal complexes [Ln (H2O) 5 (Cy3PO) 2] 3+{\textperiodcentered} Cy3PO{\textperiodcentered} 3Cl- Ln= Dy, Er},
  author={Lees, Anthony MJ and Platt, Andrew WG},
  journal={Polyhedron},
  volume={67},
  pages={368--372},
  year={2014},
  publisher={Elsevier}
}

@article{neese2020orca,
  title={The ORCA quantum chemistry program package},
  author={Neese, Frank and Wennmohs, Frank and Becker, Ute and Riplinger, Christoph},
  journal={The Journal of chemical physics},
  volume={152},
  number={22},
  pages={224108},
  year={2020},
  publisher={AIP Publishing},
  doi={10.1063/5.0004608}
}

@article{MPP,
  title = {Two Simple and Reliable Metrics of Molecular Planarity: Molecular Planarity Parameter (MPP) and Span of Deviation from Plane (SDP)},
  DOI = {10.26434/chemrxiv.14740344.v1},
  publisher = {American Chemical Society (ACS)},
  author = {Lu,  Tian},
  year = {2021},
  month = jun 
}

@Article {molSimplify,
author = {Ioannidis, Efthymios I. and Gani, Terry Z. H. and Kulik, Heather J.},
title = {mol{S}implify: A Toolkit for Automating Discovery in Inorganic Chemistry},
journal = {Journal of Computational Chemistry},
volume = {37},
number = {22},
pages = {2106--2117},
issn = {1096-987X},
doi = {10.1002/jcc.24437},
year = {2016},
}

@article{Bartok_2013,
  title = {On representing chemical environments},
  author = {Bart\'ok, Albert P. and Kondor, Risi and Cs\'anyi, G\'abor},
  journal = {Phys. Rev. B},
  volume = {87},
  issue = {18},
  pages = {184115},
  numpages = {16},
  year = {2013},
  month = {May},
  publisher = {American Physical Society},
  doi = {10.1103/PhysRevB.87.184115},
}

@article{hingorani2015review,
  title={A review of responsive MRI contrast agents: 2005--2014},
  author={Hingorani, Dina V and Bernstein, Adam S and Pagel, Mark D},
  journal={Contrast media \& molecular imaging},
  volume={10},
  number={4},
  pages={245--265},
  year={2015},
  publisher={Wiley Online Library}
}

@article{sessoli2009strategies,
  title={Strategies towards single molecule magnets based on lanthanide ions},
  author={Sessoli, Roberta and Powell, Annie K},
  journal={Coordination Chemistry Reviews},
  volume={253},
  number={19-20},
  pages={2328--2341},
  year={2009},
  publisher={Elsevier}
}

@article{coronado2020molecular,
  title={Molecular magnetism: from chemical design to spin control in molecules, materials and devices},
  author={Coronado, Eugenio},
  journal={Nature Reviews Materials},
  volume={5},
  number={2},
  pages={87--104},
  year={2020},
  publisher={Nature Publishing Group UK London}
}

@article{bogani2008molecular,
  title={Molecular spintronics using single-molecule magnets},
  author={Bogani, Lapo and Wernsdorfer, Wolfgang},
  journal={Nature materials},
  volume={7},
  number={3},
  pages={179--186},
  year={2008},
  publisher={Nature Publishing Group UK London}
}

@article{sessoli1993high,
  title={High-spin molecules:[Mn12O12 (O2CR) 16 (H2O) 4]},
  author={Sessoli, Roberta and Tsai, Hui Lien and Schake, Ann R and Wang, Sheyi and Vincent, John B and Folting, Kirsten and Gatteschi, Dante and Christou, George and Hendrickson, David N},
  journal={Journal of the American Chemical Society},
  volume={115},
  number={5},
  pages={1804--1816},
  year={1993},
  publisher={ACS Publications}
}

@article{rinehart2011exploiting,
  title={Exploiting single-ion anisotropy in the design of f-element single-molecule magnets},
  author={Rinehart, Jeffrey D and Long, Jeffrey R},
  journal={Chemical Science},
  volume={2},
  number={11},
  pages={2078--2085},
  year={2011},
  publisher={Royal Society of Chemistry}
}

@article{lunghi2022toward,
  title={Toward exact predictions of spin-phonon relaxation times: An ab initio implementation of open quantum systems theory},
  author={Lunghi, Alessandro},
  journal={Science Advances},
  volume={8},
  number={31},
  pages={eabn7880},
  year={2022},
  publisher={American Association for the Advancement of Science}
}

@article{mariano2025role,
  title={The role of electronic excited states in the spin-lattice relaxation of spin-1/2 molecules},
  author={Mariano, Lorenzo A and Nguyen, Vu Ha Anh and Petersen, Jonatan B and Bj{\"o}rnsson, Magnus and Bendix, Jesper and Eaton, Gareth R and Eaton, Sandra S and Lunghi, Alessandro},
  journal={Science Advances},
  volume={11},
  number={7},
  pages={eadr0168},
  year={2025},
  publisher={American Association for the Advancement of Science}
}

@article{tingle2023zinc,
  title={ZINC-22- A free multi-billion-scale database of tangible compounds for ligand discovery},
  author={Tingle, Benjamin I and Tang, Khanh G and Castanon, Mar and Gutierrez, John J and Khurelbaatar, Munkhzul and Dandarchuluun, Chinzorig and Moroz, Yurii S and Irwin, John J},
  journal={Journal of chemical information and modeling},
  volume={63},
  number={4},
  pages={1166--1176},
  year={2023},
  publisher={ACS Publications}
}

@article{mariano2024charting,
  title={Charting regions of cobalt’s chemical space with maximally large magnetic anisotropy: A computational high-throughput study},
  author={Mariano, Lorenzo A and Nguyen, Vu Ha Anh and Briganti, Valerio and Lunghi, Alessandro},
  journal={Journal of the American Chemical Society},
  volume={146},
  number={49},
  pages={34158--34166},
  year={2024},
  publisher={ACS Publications}
}

@article{kuhne2020cp2k,
  title={{CP2K}: An electronic structure and molecular dynamics software package-Quickstep: Efficient and accurate electronic structure calculations},
  author={K{\"u}hne, T. D. and Iannuzzi, M. and Del Ben, M. and Rybkin, V. V. and Seewald, P. and Stein, F. and Laino, T. and Khaliullin, R. Z. and Sch{\"u}tt, O. and Schiffmann, F. and D. Golze and J. Wilhelm and S. Chulkov and M. H. Bani-Hashemian and V. Weber and U. Bor\v{s}tnik and M. Taillefumier and A. S. Jakobovits and A. Lazzaro and H. Pabst and T. M\"{u}ller and R. Schade and M. Guidon and S. Andermatt and N. Holmberg and G. K. Schenter and A. Hehn and A. Bussy and F. Belleflamme and G. Tabacchi and A. Gl\{"o}\{ss}

@article{goodwin2017molecular,
  title={Molecular magnetic hysteresis at 60 kelvin in dysprosocenium},
  author={Goodwin, Conrad AP and Ortu, Fabrizio and Reta, Daniel and Chilton, Nicholas F and Mills, David P},
  journal={Nature},
  volume={548},
  number={7668},
  pages={439--442},
  year={2017},
  publisher={Nature Publishing Group UK London}
}

@article{Atomsk,
  title = {Atomsk: A tool for manipulating and converting atomic data file},
  volume = {197},
  DOI = {10.1016/j.cpc.2015.07.012},
  journal = {Computer Physics Communications},
  year = {2015},
  month = dec,
  pages = {212–219},
  author = {Hirel, P.}
}

@BOOK{Coey2010-qj,
  title     = "Magnetism and magnetic materials",
  author    = "Coey, J",
  publisher = "Cambridge University Press",
  month     =  mar,
  year      =  2010,
  address   = "Cambridge, England",
  language  = "en"
}

@article{mannini2009magnetic,
  title={Magnetic memory of a single-molecule quantum magnet wired to a gold surface},
  author={Mannini, Matteo and Pineider, Francesco and Sainctavit, Philippe and Danieli, Chiara and Otero, Edwige and Sciancalepore, Corrado and Talarico, Anna Maria and Arrio, Marie-Anne and Cornia, Andrea and Gatteschi, Dante and others},
  journal={Nature materials},
  volume={8},
  number={3},
  pages={194--197},
  year={2009},
  publisher={Nature Publishing Group UK London}
}

@article{zabala2021single,
  title={Single-Molecule Magnets: From Mn12-ac to dysprosium metallocenes, a travel in time},
  author={Zabala-Lekuona, Andoni and Seco, Jos{\'e} Manuel and Colacio, Enrique},
  journal={Coordination Chemistry Reviews},
  volume={441},
  pages={213984},
  year={2021},
  publisher={Elsevier}
}

@article{mcclain2018high,
  title={High-temperature magnetic blocking and magneto-structural correlations in a series of dysprosium (III) metallocenium single-molecule magnets},
  author={McClain, K Randall and Gould, Colin A and Chakarawet, Khetpakorn and Teat, Simon J and Groshens, Thomas J and Long, Jeffrey R and Harvey, Benjamin G},
  journal={Chemical Science},
  volume={9},
  number={45},
  pages={8492--8503},
  year={2018},
  publisher={Royal Society of Chemistry}
}

@article{guo2018magnetic,
  title={Magnetic hysteresis up to 80 kelvin in a dysprosium metallocene single-molecule magnet},
  author={Guo, Fu-Sheng and Day, Benjamin M and Chen, Yan-Cong and Tong, Ming-Liang and Mansikkam{\"a}ki, Akseli and Layfield, Richard A},
  journal={Science},
  volume={362},
  number={6421},
  pages={1400--1403},
  year={2018},
  publisher={American Association for the Advancement of Science}
}

@article{mondal2025spin,
  title={The spin-phonon relaxation mechanism of single-molecule magnets in the presence of strong exchange coupling},
  author={Mondal, Sourav and Netz, Julia and Hunger, David and Suhr, Simon and Sarkar, Biprajit and van Slageren, Joris and Köhn, Andreas and Lunghi, Alessandro},
  journal={ACS Central Science},
  volume={11},
  number={4},
  pages={550--559},
  year={2025},
  publisher={ACS Publications}
}

@article{demir2017giant,
  title={Giant coercivity and high magnetic blocking temperatures for N23- radical-bridged dilanthanide complexes upon ligand dissociation},
  author={Demir, Selvan and Gonzalez, Miguel I and Darago, Lucy E and Evans, William J and Long, Jeffrey R},
  journal={Nature Communications},
  volume={8},
  number={1},
  pages={2144},
  year={2017},
  publisher={Nature Publishing Group UK London}
}

@article{gould2022ultrahard,
  title={Ultrahard magnetism from mixed-valence dilanthanide complexes with metal-metal bonding},
  author={Gould, Colin A and McClain, K Randall and Reta, Daniel and Kragskow, Jon GC and Marchiori, David A and Lachman, Ella and Choi, Eun-Sang and Analytis, James G and Britt, R David and Chilton, Nicholas F and others},
  journal={Science},
  volume={375},
  number={6577},
  pages={198--202},
  year={2022},
  publisher={American Association for the Advancement of Science}
}

@incollection{lunghi2023spin,
  title={Spin-phonon relaxation in magnetic molecules: Theory, predictions and insights},
  author={Lunghi, Alessandro},
  booktitle={Computational Modelling of Molecular Nanomagnets},
  pages={219--289},
  year={2023},
  publisher={Springer}
}

@article{mondal2022unraveling,
  title={Unraveling the contributions to spin--lattice relaxation in kramers single-molecule magnets},
  author={Mondal, Sourav and Lunghi, Alessandro},
  journal={Journal of the American Chemical Society},
  volume={144},
  number={50},
  pages={22965--22975},
  year={2022},
  publisher={ACS Publications}
}

@article{neese2019chemistry,
  title={Chemistry and quantum mechanics in 2019: give us insight and numbers},
  author={Neese, Frank and Atanasov, Mihail and Bistoni, Giovanni and Maganas, Dimitrios and Ye, Shengfa},
  journal={Journal of the American Chemical Society},
  volume={141},
  number={7},
  pages={2814--2824},
  year={2019},
  publisher={ACS Publications}
}

@article{curtarolo2013high,
  title={The high-throughput highway to computational materials design},
  author={Curtarolo, Stefano and Hart, Gus LW and Nardelli, Marco Buongiorno and Mingo, Natalio and Sanvito, Stefano and Levy, Ohad},
  journal={Nature materials},
  volume={12},
  number={3},
  pages={191--201},
  year={2013},
  publisher={Nature Publishing Group UK London}
}

@article{moosavi2020role,
  title={The role of machine learning in the understanding and design of materials},
  author={Moosavi, Seyed Mohamad and Jablonka, Kevin Maik and Smit, Berend},
  journal={Journal of the American Chemical Society},
  volume={142},
  number={48},
  pages={20273--20287},
  year={2020},
  publisher={ACS Publications}
}

@article{sanchez2018inverse,
  title={Inverse molecular design using machine learning: Generative models for matter engineering},
  author={Sanchez-Lengeling, Benjamin and Aspuru-Guzik, Al{\'a}n},
  journal={Science},
  volume={361},
  number={6400},
  pages={360--365},
  year={2018},
  publisher={American Association for the Advancement of Science}
}

@article{lunghi2022computational,
  title={Computational design of magnetic molecules and their environment using quantum chemistry, machine learning and multiscale simulations},
  author={Lunghi, Alessandro and Sanvito, Stefano},
  journal={Nature Reviews Chemistry},
  volume={6},
  number={11},
  pages={761--781},
  year={2022},
  publisher={Nature Publishing Group UK London}
}

@article{alemany2017continuous,
  title={Continuous symmetry measures: a new tool in quantum chemistry},
  author={Alemany, Pere and Casanova, David and Alvarez, Santiago and Dryzun, Chaim and Avnir, David},
  journal={Reviews in Computational Chemistry},
  volume={30},
  pages={289--352},
  year={2017},
  publisher={Wiley Online Library}
}

@article{dobson2004chemical,
  title={Chemical space and biology.},
  author={Dobson, Christopher M},
  journal={Nature},
  volume={432},
  number={7019},
  year={2004}
}

@article{price2014predicting,
  title={Predicting crystal structures of organic compounds},
  author={Price, Sarah L},
  journal={Chemical Society Reviews},
  volume={43},
  number={7},
  pages={2098--2111},
  year={2014},
  publisher={Royal Society of Chemistry}
}

@article{thakkar2021retrosynthetic,
  title={Retrosynthetic accessibility score (RAscore)--rapid machine learned synthesizability classification from AI driven retrosynthetic planning},
  author={Thakkar, Amol and Chadimov{\'a}, Veronika and Bjerrum, Esben Jannik and Engkvist, Ola and Reymond, Jean-Louis},
  journal={Chemical science},
  volume={12},
  number={9},
  pages={3339--3349},
  year={2021},
  publisher={Royal Society of Chemistry}
}

@article{ertl2009estimation,
  title={Estimation of synthetic accessibility score of drug-like molecules based on molecular complexity and fragment contributions},
  author={Ertl, Peter and Schuffenhauer, Ansgar},
  journal={Journal of cheminformatics},
  volume={1},
  number={1},
  pages={8},
  year={2009},
  publisher={Springer}
}

@article{matta2003hydrogen,
  title={Hydrogen--hydrogen bonding: a stabilizing interaction in molecules and crystals},
  author={Matta, Ch{\'e}rif F and Hern{\'a}ndez-Trujillo, Jes{\'u}s and Tang, Ting-Hua and Bader, Richard FW},
  journal={Chemistry--A European Journal},
  volume={9},
  number={9},
  pages={1940--1951},
  year={2003},
  publisher={Wiley Online Library}
}

@article{matta2006extended,
  title={Extended weak bonding interactions in DNA: $\pi$-stacking (base- base), base- backbone, and backbone- backbone interactions},
  author={Matta, Ch{\'e}rif F and Castillo, Norberto and Boyd, Russell J},
  journal={The Journal of Physical Chemistry B},
  volume={110},
  number={1},
  pages={563--578},
  year={2006},
  publisher={ACS Publications}
}

@article{frangoulis2025generating,
  title={Generating New Coordination Compounds via Multireference Simulations, Genetic Algorithms, and Machine Learning: The Case of Co (II) and Dy (III) Molecular Magnets},
  author={Frangoulis, Lion and Khatibi, Zahra and Mariano, Lorenzo A and Lunghi, Alessandro},
  journal={JACS Au},
  volume={5},
  number={8},
  pages={3808--3821},
  year={2025},
  publisher={ACS Publications}
}

@incollection{eaton2000relaxation,
  title={Relaxation times of organic radicals and transition metal ions},
  author={Eaton, Sandra S and Eaton, Gareth R},
   editor={Berliner, Lawrence J and Eaton, Sandra S and Eaton, Gareth R},
  booktitle={Distance measurements in biological systems by EPR},
  pages={29--154},
  year={2000},
  publisher={Springer US}
}

@article{mcadams2017molecular,
  title={Molecular single-ion magnets based on lanthanides and actinides: Design considerations and new advances in the context of quantum technologies},
  author={McAdams, Simon G and Ariciu, Ana-Maria and Kostopoulos, Andreas K and Walsh, James PS and Tuna, Floriana},
  journal={Coordination Chemistry Reviews},
  volume={346},
  pages={216--239},
  year={2017},
  publisher={Elsevier}
}

@article{choudhary2022recent,
  title={Recent advances and applications of deep learning methods in materials science},
  author={Choudhary, Kamal and DeCost, Brian and Chen, Chi and Jain, Anubhav and Tavazza, Francesca and Cohn, Ryan and Park, Cheol Woo and Choudhary, Alok and Agrawal, Ankit and Billinge, Simon JL and others},
  journal={npj Computational Materials},
  volume={8},
  number={1},
  pages={59},
  year={2022},
  publisher={Nature Publishing Group UK London}
}

@article{luo2025supramolecular,
 author = {Qian-Cheng Luo  and Lorenzo A. Mariano  and Wen-Jie Xu  and Zi-Han Li  and Ke-Xin Yu  and Yan Peng  and Alessandro Lunghi  and Yan-Zhen Zheng },
title = {Supramolecular Interactions Modulate Raman Relaxation in D5h Symmetric Dy(III) Single-Molecule Magnets Opening the Magnetic Hysteresis up to 50 K},
journal = {ChemRxiv},
volume = {2025},
number = {0731},
pages = {},
year = {2025},
}

@article{swain2025encapsulation,
  title={Encapsulation Enhances the Quantum Coherence of a Solid-State Molecular Spin Qubit},
  author={Swain, Abinash and Barrios, Leon{\'\i} A and Nelyubina, Yulia and Teat, Simon J and Roubeau, Olivier and Novikov, Valentin and Arom{\'\i}, Guillem},
  journal={Angewandte Chemie},
  volume={137},
  number={42},
  pages={e202510603},
  year={2025},
  publisher={Wiley Online Library}
}

\end{document}